%% file: main.tex
\documentclass[conference]{IEEEtran}
\IEEEoverridecommandlockouts
\usepackage{amsmath,amssymb,amsfonts}
\usepackage{algorithmic}
\usepackage{graphicx}
\usepackage{textcomp}
\usepackage{xcolor}
\def\BibTeX{{\rm B\kern-.05em{\sc i\kern-.025em b}\kern-.08em
    T\kern-.1667em\lower.7ex\hbox{E}\kern-.125emX}}
\usepackage{pgfplots}
\pgfplotsset{compat=1.18}
\usetikzlibrary{positioning}
\usepgfplotslibrary{external}
\usepgfplotslibrary{statistics}
\usepackage{caption}
\usepackage{subcaption}
\usepackage{csquotes}
\usepackage{url}
\usepackage{multirow}
\usepackage{graphics}
\usepackage{enumitem}
\usepackage{color, colortbl}
\usepackage{lipsum}
\usepackage{tabularx,booktabs}
\usepackage{pdflscape}
\usepackage{adjustbox}
\usepackage{tcolorbox}
\usepackage{array}
\usepackage{ragged2e}
\usepackage{afterpage} 
\usepackage{rotating}
\usepackage{graphicx}
\usepackage{comment}
\usepackage[para]{footmisc}
\usepackage{tablefootnote}
\usepackage[nobiblatex]{xurl}
\definecolor{lightgray}{rgb}{0.83, 0.83, 0.83}
\DeclareUnicodeCharacter{202F}{FIX ME!!!!}

\renewcommand\thefigure{\arabic{figure}}
\DeclareCaptionLabelFormat{simplab}{Figure~#2}
\DeclareCaptionFormat{nocap}{}
\DeclareCaptionSubType*{figure}

\makeatletter
\def\@copyrightspace{\relax}
\makeatother

\makeatletter
\newcommand\notsotiny{\@setfontsize\notsotiny{5.5}{6.5}}
\makeatother

\title{Infrastructure Patterns in Toll Scam Domains: A Comprehensive Analysis of Cybercriminal Registration and Hosting Strategies}

\author{\IEEEauthorblockN{Morium Akter Munny}
\IEEEauthorblockA{California State University San Marcos\\
munny001@csusm.edu}
\and
\IEEEauthorblockN{Mahbub Alam}
\IEEEauthorblockA{Texas A\&M University\\
mahbub.alam@tamu.edu}
\and
\IEEEauthorblockN{Sonjoy Kumar Paul}
\IEEEauthorblockA{Texas A\&M University\\
skpaul@tamu.edu}
\and
\IEEEauthorblockN{Daniel Timko}
\IEEEauthorblockA{Emerging Threats Lab / Smishtank.com\\
daniel@smishtank.com}
\and
\IEEEauthorblockN{Muhammad Lutfor Rahman}
\IEEEauthorblockA{California State University San Marcos\\
mlrahman@csusm.edu}
\and
\IEEEauthorblockN{Nitesh Saxena}
\IEEEauthorblockA{Texas A\&M University\\
nsaxena@tamu.edu}}

\begin{document}
\bstctlcite{bstctl:etal, bstctl:nodash, bstctl:simpurl}
\maketitle

\begin{abstract}
\input{abstract.tex}
\end{abstract}


\input{introduction}
\input{background}
\input{relatedworks}
\input{methodology}

\input{results}
\input{discussion}
\input{conclusion}
\input{acknowledgment}

\bibliographystyle{IEEEtran}
\bibliography{toll-scam}
\end{document}

%% file: abstract.tex
Toll scams involve criminals registering fake domains that pretend to be legitimate transportation agencies to trick users into making fraudulent payments. Although these scams are rapidly increasing and causing significant harm, they have not been extensively studied. We present the first large-scale analysis of toll scam domains, using a newly created dataset of 67,907 confirmed scam domains mostly registered in 2025. Our study reveals that attackers exploit permissive registrars and less common top-level domains, with 86.9\% of domains concentrated in just five non-mainstream TLDs and 72.9\% registered via a single provider. We also discover specific registration patterns, including short bursts of activity that suggest automated, coordinated attacks, with over half of domains registered in the first quarter of 2025. This extreme temporal clustering reflects highly synchronized campaign launches. Additionally, we build a simple predictive model using only domain registration data to predict which scam domains are likely to be suspended---a proxy for confirmed abuse---achieving 80.4\% accuracy, and 92.3\% sensitivity. Our analysis reveals attacker strategies for evading detection---such as exploiting obscure TLDs, permissive registrars, and coordinated registration bursts---which can inform more targeted interventions by registrars, hosting providers, and security platforms. However, our results suggest that registration metadata alone may be insufficient, and incorporating features from domain URLs and webpage content could further improve detection.

%% file: introduction.tex
\section{Introduction}

Toll scamming is a newly emerging cybercrime tactic that has rapidly gained prominence~\cite{mineo2024youd, rayo2025got}. In April 2024, the FBI's Internet Crime Complaint Center (IC3) issued a public service announcement explicitly warning consumers about fraudulent toll service messages impersonating legitimate transportation agencies and coercing victims into visiting malicious websites for fake fine payments~\cite{IC3-2024-PSA240412}. According to the FBI’s 2024 IC3 report~\cite{fbi_ic3_annualreport2024}, over 59,000 complaints were linked to toll scams, signaling both their rapid growth and broad societal impact. While individual losses are often small (e.g., \$5–\$10), these attacks can have serious consequences, including credit card theft and downstream identity fraud---amplifying their impact despite the modest per-incident cost~\cite{viu2025cybercrime}. Moreover, the low financial stakes reduce the likelihood that victims will report the incident to authorities, allowing many such scams to go untracked and enabling attackers to operate at scale with minimal disruption~\cite{bidgoli2017hello}.

Toll scams occupy a unique position in the phishing landscape. Like spear phishing, they appear personalized---using toll authority names tailored to the victim's state or region, inferred from their phone number area code~\cite{Gibson2025UnpaidTollScam}. However, unlike traditional spear phishing, this form of targeting does not require detailed personal information, making it inexpensive and easily scalable. This lightweight personalization increases the perceived legitimacy of the message, making recipients more likely to trust and act on it~\cite{hable2025phishing}.

Despite these characteristics, toll scams remain largely absent from cybersecurity literature. Prior work on domain-based abuse has focused primarily on phishing~\cite{hao_predator_2016}, combosquatting~\cite{kintis_hiding_2017}, and malware distribution~\cite{coull_understanding_2012}, with limited attention to the infrastructure and registration behaviors unique to toll scam operations. Key questions remain unaddressed: Do these campaigns rely on specific top-level domains (TLDs), registrar patterns, or burst-based registrations? Do they exhibit evasive behaviors that complicate early detection?

Detecting toll scam domains poses inherent challenges due to their adaptive and deceptive nature. Attackers often imitate trusted tolling authorities~\cite{rayo2025got}, use lookalike or typosquatting domains~\cite{agten2015seven}, and exploit obscure or under-regulated TLDs to avoid detection~\cite{lim_registration_2025}. Addressing this threat requires a deeper understanding of attacker behavior and infrastructure patterns.

To fill this gap, we introduce the first large-scale dataset of 67,907 confirmed toll scam domains, primarily registered in 2025. This dataset was provided by the State of Oklahoma, which manually verified scam domains based on user reports and internal investigations. Leveraging this ground-truth dataset, we conduct the first in-depth characterization of toll scam infrastructure, registration practices, and temporal activity. Finally, we assess whether simple registration metadata---without relying on domain content, lexical features, or behavioral signals---can support predictive detection of toll scam domains at registration time. Our study is guided by the following research questions:

\begin{itemize}
    \item \textbf{RQ1}: What infrastructure patterns distinguish suspended from active toll scam domains?

    \item \textbf{RQ2}: How do toll scam registration patterns vary across different TLDs and registrars?

    \item \textbf{RQ3}: What temporal patterns exist in toll scam domain registrations?

    \item \textbf{RQ4}: Can we predict which toll scam domains are likely to be suspended based on registration metadata?

\end{itemize}

In addressing these questions, our study makes the following core contributions:

\begin{itemize}
    \item \textbf{New Dataset on Toll Scams}: We curate and release the first known dataset of 67,907 verified toll scam domains, including suspension status, registration metadata, hosting infrastructure, and temporal activity. This dataset reflects real-world abuse at scale and serves as a critical resource for future research on scam detection and infrastructure takedown.

    \item \textbf{Infrastructure Abuse Analysis}: We systematically analyze the hosting infrastructure of scam domains and find that suspended domains are highly concentrated on a few high-risk  Autonomous System Numbers (ASNs) and show much higher IP clustering. This indicates attackers often register and host large numbers of scam domains on the same networks, enabling large-scale operations but increasing susceptibility to bulk suspension.

    \item \textbf{Registration Ecosystem Mapping}: Our study reveals extreme concentration of registrations in a handful of TLDs and registrars---86.9\% of scam domains are concentrated in just five non-mainstream TLDs, and 72.9\% are registered through a single provider---exposing weak points in the domain registration supply chain and enabling insights for targeted intervention.

     \item \textbf{Temporal Pattern Discovery}: We uncover burst-based registration behaviors and identify strong temporal concentration, with more than half of all scam domains registered in the first quarter of 2025---evidence of large-scale, highly coordinated campaigns likely backed by automation.

     \item \textbf{Predictive Modeling}: We train and evaluate a lightweight machine learning model using only registration metadata (registrar, TLD, name server, registration timing) to predict which toll scam domains are likely to be suspended. Based on ICANN’s DNS Abuse Compliance Reports~\cite{ICANNReports20241224}, such suspension actions indicate that actionable evidence of abuse was identified, making suspension a meaningful operational signal for high-risk domains. Despite limited features, our model achieves 80.4\% overall accuracy and 92.3\% sensitivity, indicating that registration metadata can serve as an early warning signal for high-risk scam domains.

\end{itemize}

Overall, these findings highlight concrete opportunities for proactive intervention by registrars and security platforms, while underscoring the need for richer metadata and multi-faceted features to enable more robust scam detection.

%% file: background.tex
\section{Background}
\label{sec:background}

\begin{figure*}[!ht]
\centering
\includegraphics[scale=0.5, trim=75 130 75 120, clip]{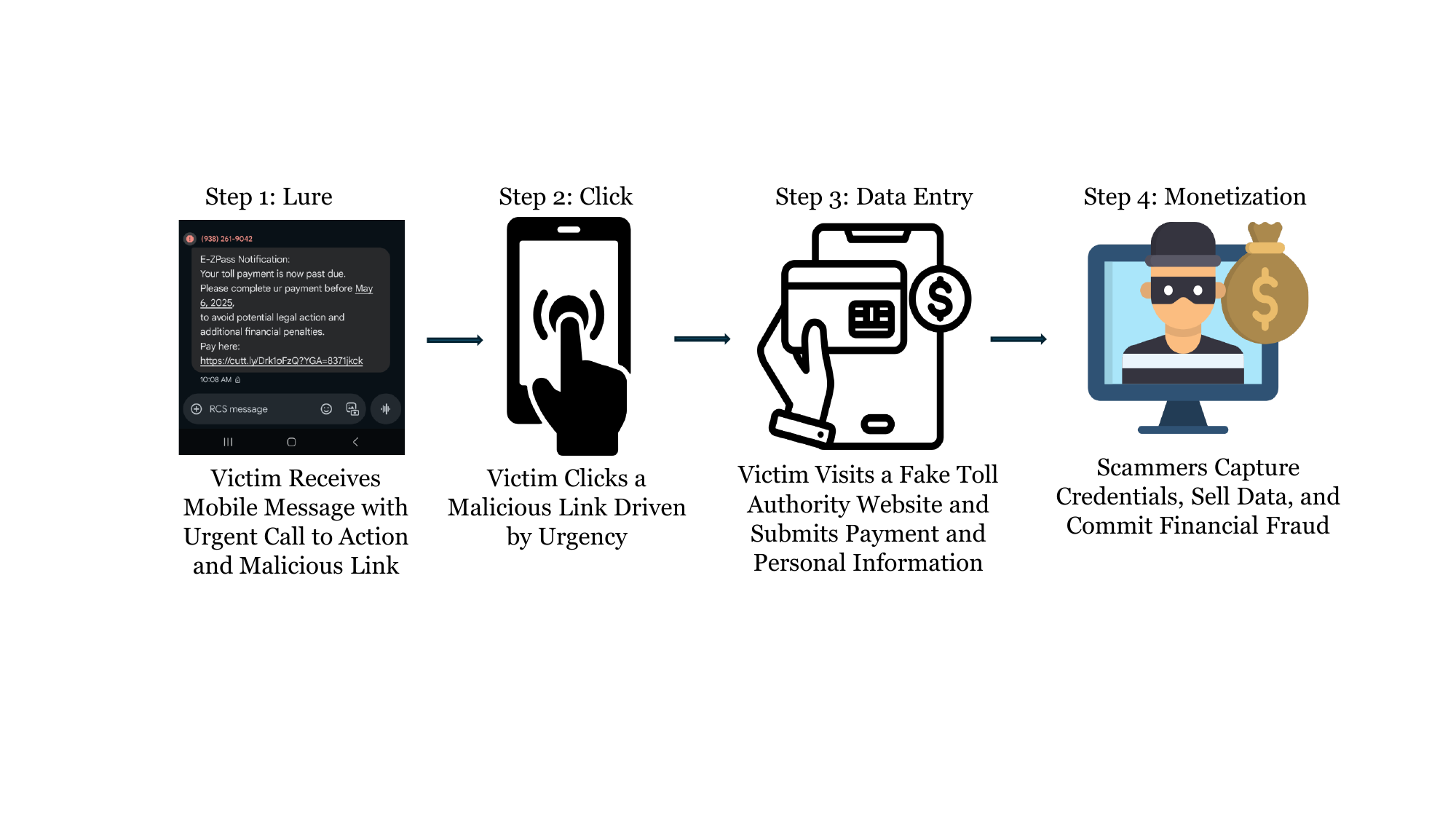}
\caption{The Toll Scam Attack Life Cycle}
\label{fig:scam_lifecycle}
\end{figure*}

The landscape of cybercrime is continually evolving, with adversaries shifting from broad, indiscriminate attacks to more targeted and contextually aware social engineering schemes \cite{alkhalil_phishing_2021}. A prominent and recent example of this trend is the rise of toll road phishing scams, a specialized form of scam texting \cite{fcc_how_2025}. These campaigns exploit the daily routines of millions of drivers, combining psychological manipulation with a rapidly changing technical infrastructure to harvest sensitive personal and financial data. This section provides background on the mechanics of these scams, the psychological principles that make them effective, and the key technical domain-related concepts central to our investigation.

\subsection{Anatomy of a Toll Scam}

Toll road smishing scams follow a clear, multi-stage process designed to guide a victim from initial contact to financial exploitation. The typical attack life-cycle illustrated in Figure \ref{fig:scam_lifecycle}, is as follows:

\begin{enumerate}
    \item \textbf{Mobile Message (Lure):} The victim receives an unsolicited message via mobile platforms (e.g., ``SMS", ``iMessage", and ``RCS"). The message claims to be from a legitimate toll authority (e.g., ``E-ZPass", ``SunPass", ``TxTag") and states that the recipient has an outstanding toll balance. To create a sense of urgency, the message often threatens an impending late fee or other penalty if the balance is not settled immediately. The message contains a URL that the victim is urged to click to resolve the payment. These URLs are often disguised using URL shorteners or by creating ``combosquatting'' domains that mimic the real authority's name (e.g., \texttt{thetollroads-paytoll.world} instead of \texttt{thetollroads.com}) \cite{kintis_hiding_2017}.

    \item \textbf{The Phishing Website (Click):} As the victim clicks the malicious link driven by urgency, the link directs the victim to a fraudulent website meticulously designed to be a pixel-perfect replica of the official toll payment portal. This page is hosted on infrastructure specifically chosen for the campaign.

    \item \textbf{Information \& Credential Harvesting (Data Entry):} The fake portal prompts the user to enter personally identifiable information (PII) such as their full name, address, and driver's license number, as well as financial details like credit card numbers, expiration dates, and CVV codes.

    \item \textbf{Monetization:} The cybercriminals immediately collect this information. The financial data is used for fraudulent purchases or sold on dark web marketplaces. The PII can be used for broader identity theft. The domain used in the attack is often quickly taken down or abandoned to evade detection and blacklisting.
\end{enumerate}

\subsection{Psychological Vulnerabilities and Uniqueness}

The success of toll scams is rooted in their exploitation of several cognitive biases and the unique context of mobile communication. Unlike generic phishing emails, these scams are effective for specific reasons:

\begin{itemize}
    \item \textbf{Authority and Urgency:} Scammers impersonate a government or quasi-governmental agency, leveraging the public's inherent tendency to comply with figures of authority. The threat of fines or legal action triggers a fear response, compelling victims to act quickly without critical thought \cite{phishing_and_Countermeasures_2006}.

    \item \textbf{Plausibility and Context:} Toll roads are a common part of daily life for many people. A driver may have recently used a toll road, making the message seem highly plausible. This creates a confirmation bias, where the victim's brain seeks to confirm a believable premise rather than challenge it \cite{dhamija_why_2006}.

    \item \textbf{Mobile-First Deception:} The attacks are delivered via message to mobile devices. Smaller screens can make it difficult to inspect the full URL of a link, and users are often distracted or multitasking when using their phones, leading to reduced scrutiny \cite{rahman2023users}. 

    \item \textbf{Uniqueness from Traditional Phishing:} The primary uniqueness lies in its hyper-contextual and targeted nature \cite{jagatic_social_2007}. While traditional phishing often uses broad lures like ``Your account has been suspended,'' toll scams tap into a specific, common, and often anxiety-inducing real-world event (paying for government services). This specificity, combined with the ephemeral nature of the infrastructure used, makes it a particularly challenging modern threat \cite{oest_sunrise_2020}.
\end{itemize}

\subsection{Key Domain and Infrastructure Concepts}

Analyzing the infrastructure behind these scams requires understanding several key technical components. The choice of these components is not random but reflects the operational security and cost considerations of the attackers.

\begin{itemize}
    \item \textbf{Domain Status:} This refers to the state of a domain name in the Domain Name System (DNS). For this study, the two primary statuses are \textbf{Active} (the domain resolves to an IP address and can host content) and \textbf{Suspended} (the domain has been deactivated by the registrar or registry, often due to abuse complaints, and no longer resolves). Analyzing the differences between these groups is the focus of our RQ1.

    \item \textbf{TLD:} This is the suffix at the end of a domain name. TLDs are managed by registries and are broadly categorized as generic TLDs (gTLDs like \texttt{.com}, \texttt{.org}), country-code TLDs (ccTLDs like \texttt{.us}, \texttt{.cn}), and newer gTLDs (e.g., \texttt{.xyz}, \texttt{.top}, \texttt{.vip}). Research shows that abusers often flock to newer, cheaper TLDs with lax registration policies \cite{korczynski_cybercrime_2018}.

    \item \textbf{Registrar:} A commercial entity accredited by ICANN and TLD registries to sell domain names to the public (the registrants). Registrars are a critical control point, as their policies and responsiveness to abuse complaints can either facilitate or hinder malicious campaigns \cite{akiwate_risky_2021}.

    \item \textbf{Hosting Infrastructure, ISP, and ASN:} The \textit{hosting infrastructure} refers to the network of servers where the fraudulent website's files are stored. Each server has an IP address, which is managed by an \textbf{Internet Service Provider (ISP)}. ISPs, in turn, are part of larger networks called \textbf{Autonomous Systems (AS)}, each identified by a unique \textbf{ASN}. Analyzing the ASNs where scam domains are hosted reveals the network operators (e.g., Amazon Web Services, Cloudflare, or more obscure bulletproof providers) that are either wittingly or unwittingly supporting the fraudulent activity \cite{konte_dynamics_2009}. These features are foundational to proactive detection systems like PREDATOR \cite{hao_predator_2016}.
\end{itemize}

%% file: relatedworks.tex
\section{Related Works}
\label{sec:relatedworks}

The study of cybercriminal infrastructure is foundational to understanding and combating online fraud. A significant body of research has investigated how malicious actors exploit the DNS, with a particular focus on phishing, malware distribution, and various scams \cite{cartron_dangers_2025}. This section provides the methodological and conceptual groundwork for investigating toll scam domains. We structure this section around four key themes that directly inform our research questions: the infrastructure and lifecycle of malicious domains, the role of the broader DNS ecosystem, the temporal dynamics of fraudulent registrations, and the predictive modeling of domain abuse using registration metadata.

\textbf{Infrastructure and Life Cycle of Malicious Domains:}
A primary focus of prior work has been to characterize the infrastructure that supports malicious online activities. Research has extensively documented the tactics used by cybercriminals to maintain operational resilience and evade detection. Foundational studies identified the use of techniques like fast-flux service networks, where IP addresses associated with a domain rapidly change, to create a moving target for takedown efforts \cite{holz_measuring_2008, caglayan_real-time_2009}. Konte et al. \cite{konte_dynamics_2009} further explored the dynamics of online scam hosting, emphasizing the role of hosting providers that are complicit in and provide stable infrastructure for long-term malicious campaigns. These tactics are directly relevant to our first research question (RQ1), which seeks to differentiate the infrastructure of active versus suspended toll scam domains.

Recent studies have provided a more granular view of the entire lifecycle of malicious domains. Oest et al. \cite{oest_sunrise_2020} conducted a large-scale analysis of phishing attacks from ``sunrise to sunset'' detailing the stages from infrastructure setup to monetization and eventual takedown. Similarly, Lim et al. \cite{lim_registration_2025}, in a study of over 690,000 phishing domains, revealed significant differences in the lifespan and hosting infrastructure of active versus suspended domains, underscoring how infrastructure choices impact persistence. Adding another layer of complexity, researchers distinguish between domains that are maliciously registered from the outset and legitimate domains that are compromised and repurposed for abuse \cite{dingledine_evil_2009, maroofi_comar_2020}. Erfan et al. \cite{erfan_owned_2024} found that a substantial portion of malicious activity originates from rented or compromised infrastructure, a factor that could influence the patterns we observe in toll scams.

Given that toll scams are often propagated via SMS, the work of Nahapetyan et al. \cite{nahapetyan_sms_2024} is particularly salient. Their deep dive into smishing tactics and infrastructure highlights the specific hosting and redirection strategies used in mobile-targeted campaigns, offering a direct parallel for our investigation. The effectiveness of infrastructure analysis depends critically on access to fresh, validated smishing datasets. Timko et al.~\cite{timko2023commercial,timko2024smishing} introduced collaborative approaches for collecting and validating SMS phishing datasets through Smishtank.com~\cite{smishtank2025}, which highlights the importance of collecting domain infrastructure metadata while attacks are active to archive attacker efforts to evade detection. This infrastructure-focused research suggests that infrastructure choices are not random but are instead indicative of an adversary's operational strategy, forming a strong basis for our hypothesis that active and suspended toll scam domains will exhibit distinct infrastructural footprints.

\textbf{The Role of TLDs and Registrars in Domain Abuse:}
The ecosystem of domain registration, involving registrars and TLDs, is a critical battleground for DNS abuse. Our second research question (RQ2) investigates how toll scam registration patterns vary across these entities. Prior research has firmly established that malicious registrations are not uniformly distributed. Korczynski et al. \cite{korczynski_cybercrime_2018} provided statistical evidence that new gTLDs suffer from a disproportionately high concentration of DNS abuse, often due to permissive registration policies or weak oversight. More recently, Moura et al. \cite{moura_characterizing_2024} characterized phishing at the scale of a ccTLD, demonstrating that abuse patterns and mitigation effectiveness can vary significantly by TLD type.

The role of registrars is equally critical. Akiwate et al. \cite{akiwate_risky_2021} demonstrated how risks can be derived directly from a registrar's name management practices, showing that certain registrars are disproportionately associated with malicious domains. This aligns with ICANN reports, which have consistently shown discrepancies in abuse levels across different registrars \cite{ICANNReports20171975}. Some studies have focused on the malicious ecosystem within a specific TLD, such as the .eu TLD, mapping the relationships between registrants and registrars involved in abuse \cite{vissers_exploring_2017}. The methodologies for measuring this abuse are also a key topic of research, with Tajalizadehkhoob et al. \cite{tajalizadehkhoob_rotten_2018} calling for more robust statistical methods to differentiate systemic issues at a registrar or TLD from isolated instances of abuse. This wealth of research confirms that TLD and registrar data are strong features for characterizing malicious activity, directly supporting the premise of RQ2.

\textbf{Temporal Dynamics of Malicious Registrations:}
Cybercriminal campaigns are often characterized by distinct temporal patterns, a phenomenon we investigate for toll scams in our third research question (RQ3). Early work by Coull et al. \cite{coull_understanding_2012} identified abnormal bulk domain registrations as a key indicator of malicious intent, linking temporal spikes in registration activity to domaining abuses like tasting and speculation. This focus on registration behavior was expanded by Hao et al. \cite{hao_understanding_2013}, who analyzed the specific registration patterns of spammers.

More recently, large-scale longitudinal studies have confirmed that temporal clustering is a hallmark of coordinated campaigns. Agarwal and Vasek \cite{agarwal_examining_2025}, in their examination of newly registered phishing domains, identified significant temporal bunching, suggesting synchronized campaign launches. The lifecycle analysis by Oest et al. \cite{oest_sunrise_2020} also contributes to this understanding by mapping the timeline of attacker activities. The initial moments of a domain's life are particularly revealing; Hao et al. \cite{hao_monitoring_2011} showed that monitoring the initial DNS behavior of a domain can be highly indicative of its future malicious use. These studies provide a strong precedent for our temporal analysis, suggesting that examining the timing and clustering of toll scam domain registrations can reveal evidence of coordinated fraudulent campaigns.

\textbf{Predictive Modeling Using Registration Metadata:}
Our final research question (RQ4) explores the potential for predicting toll scam domains using their registration metadata. This goal is well-supported by a robust history of research applying data analytics and machine learning to this problem. A comprehensive survey by Zhauniarovich et al. \cite{zhauniarovich_survey_2018} outlines the vast landscape of DNS data analysis for malicious domain detection.

One of the most successful approaches has been proactive detection at the time of registration. Hao et al. \cite{hao_predator_2016} developed PREDATOR, a system that successfully identified malicious domains by leveraging features available at registration, including bulk registration patterns, registrant information, and infrastructure reuse. This work demonstrated the predictive power of metadata before a domain is even used in an attack. Subsequent research has refined this approach, using machine learning to detect specific abuses like typo-squatting and combosquatting based on domain naming patterns and registration data \cite{moubayed_dns_2018, kintis_hiding_2017, shirazi_kn0w_2018}. Systems like Phishnet have shown the viability of predictive blacklisting based on these features \cite{prakash_phishnet_2010}. These studies establish that metadata such as the choice of registrar, TLD, ASN, and lexical features of the domain name are highly predictive of malicious intent, providing a strong methodological foundation for our predictive modeling efforts in RQ4.

In summary, the existing literature provides a rich foundation for our study. It establishes the importance of analyzing infrastructure, the DNS ecosystem, and temporal patterns to understand and combat online fraud. However, while these principles have been extensively applied to phishing, malware, and spam, no prior work has conducted a comprehensive, data-driven investigation into the unique characteristics of toll scam domains. Our research aims to fill this critical gap, applying established methodologies to a novel and growing threat.

%% file: methodology.tex
\section{Methodology}
\label{sec:methodology}

This section outlines the datasets used in our research, the enrichment process we applied to domain-level metadata, and the analytical methods employed to address our research questions (RQ1–RQ4).

\subsection{Datasets}
We compiled a dataset of 67,907 confirmed toll scam domains, provided by the State of Oklahoma. These domains were manually verified through user reports and internal investigations. We will release the dataset to support future research on domain abuse and toll scam detection.

These domains impersonate legitimate toll authorities and payment portals to trick users into paying fraudulent toll fees or violation fines.

\subsection{Data Preprocessing and Enrichment}

To support infrastructure-level and predictive analysis, we performed several data cleaning and enrichment steps:

\begin{itemize}
    \item\textbf{Registrar Standardization:} Registrar names often appeared in multiple syntactic variants (e.g. ``NameSilo, LLC'', ``NameSilo: 36''), which were normalized to unified labels. This consolidation ensured consistent counting and analysis of registrar behavior across the dataset.
    \textbf{TLD Normalization:} TLDs were reviewed for typos, formatting inconsistencies, and aliases. After cleaning, low-frequency TLDs were grouped under a category ``Other'' to mitigate the effects of the sparse matrix in categorical modeling.

\item \textbf{Domain Status Recoding:} The original domain status field was transformed into a binary variable to facilitate inferential tests and classification. Domains labeled as ``clientHold'' or ``serverHold'' were recoded as \texttt{1} (Suspended), while all other statuses were recoded as \texttt{0} (Active).

\item \textbf{IP Clustering Feature Generation:} A clustering metric was introduced to capture the extent to which multiple domains share the same IP address. This variable, denoted as {ip\_domain\_count}, was computed by aggregating the number of domains registered under each unique IP. High clustering values may indicate shared hosting infrastructure used for bulk scam operations.

\item \textbf{Temporal Feature Extraction:} From the WHOIS creation date, we extracted structured temporal indicators including \texttt{year}, \texttt{month}, and a combined \texttt{year\_month} field. These features supported longitudinal and seasonal analyses of scam registration activity.

\item \textbf{ASN and Registrar Grouping:} To reduce noise and improve the generalizability of the model, ASNs and Registry entities with very low representation were grouped into the ``Other'' category. This step preserved dominant patterns while preventing overfitting in categorical analyses.

\end{itemize}

Collectively, these enrichment techniques improved the analytical utility of the dataset by improving consistency, reducing sparsity, and introducing new dimensions of knowledge. The resulting features formed the basis for subsequent statistical analyses and classification tasks in this study.

\subsection{Statistical Analysis}

To address each research question (RQ), we applied a combination of statistical and machine learning techniques tailored to the structure and goals of each analysis:

\begin{itemize}

    \item \textbf{RQ1 – Infrastructure–Status Relationship:} We used Pearson's Chi-square test to evaluate the association between domain status (suspended vs. active) and ASNs. This test is well-suited for detecting categorical dependencies and is widely used in cybersecurity infrastructure studies~\cite{akiwate_risky_2021}. To compare IP clustering between suspended and active domains, we applied both an independent-sample $t$-test and the non-parametric Mann--Whitney $U$ test. While the $t$-test assesses mean differences under the assumption of normality, the Mann--Whitney $U$ test provides a robust alternative that compares the distributions without assuming normality---an important consideration given the typically skewed nature of domain-to-IP mappings. Reporting both tests ensures that our findings are not artifacts of distributional assumptions and remain valid under real-world conditions.

    \item \textbf{RQ2 – TLD and Registrar Usage Patterns:} We conducted frequency analysis to identify which TLDs and registrars were most commonly used by toll scam operators, following established methods in cybersecurity research~\cite{korczynski_cybercrime_2018}. We initially planned to use chi-square tests of independence to examine whether TLD and registrar choices are systematically related, but the extreme concentration of registrations in a few TLDs and registrars resulted in insufficient variation for reliable statistical testing~\cite{akiwate_risky_2021}. This concentration pattern itself represents a significant finding about systematic infrastructure exploitation. Instead, we applied $K$-Means clustering, a method widely used in cybersecurity research for identifying malicious domain patterns, to automatically group similar domains based on their registration characteristics~\cite{vissers_exploring_2017,zhauniarovich_survey_2018}. This clustering approach can discover hidden patterns in criminal operations, such as coordinated campaigns that use similar registration strategies, without requiring prior knowledge of campaign structures.

    \item \textbf{RQ3 -- Temporal Registration Patterns:} To identify when toll scam domains are registered and whether they follow specific timing patterns, we used several statistical methods that distinguish between random registrations and coordinated criminal campaigns. We organized registration dates by year and month, then counted how many domains were registered each month, creating a timeline that shows when registration activity peaks and whether domains cluster during certain periods, following approaches used in domain registration behavior analysis~\cite{hao_understanding_2013}.

The \textit{Runs Test} examines whether monthly registration patterns happen randomly or in organized bursts by detecting non-random sequences in the data---this helps determine if criminals coordinate their domain registrations or act independently~\cite{field_discovering_2018}. \textit{Autocorrelation analysis} (ACF and PACF) checks whether high registration activity in one month affects activity in following months, helping us understand if campaigns build momentum over time or operate as separate, isolated events~\cite{chatfield_analysis_2019}. \textit{ANOVA} with \textit{Tukey HSD post hoc} comparisons tests whether certain months have significantly more domain registrations than others, identifying the specific timing of coordinated criminal activities~\cite{field_discovering_2018}.

These complementary methods provide clear evidence about how toll-scam operators coordinate their timing, showing whether domains are registered steadily by individual criminals or systematically during planned campaign periods.

\item \textbf{RQ4 – Predictive Modeling of Scam Risk:} To determine whether registration metadata can predict which toll scam domains are likely to be suspended, we developed binary classification models using infrastructure patterns from RQ2. We created risk indicators for TLD and registrar data, where high-risk categories were defined based on concentration patterns: five TLDs (.xin, .top, .vip, .world, .win) accounting for 86.9\% of registrations, and three registrars (Dominet HK Limited, NameSilo, Gname) handling 91.8\% of registrations.

We then used \textit{binary logistic regression}, which helps us to understand how much each factor affects the likelihood that a domain will be suspended by using odds ratios~\cite{field_discovering_2018}. To make sure our model was accurate, we ran the \textit{Hosmer--Lemeshow test}, which checks if the model's predictions match what actually happens. We also measured how well the model could tell suspended domains apart from active ones using an \textit{ROC curve} and the AUC score, which is a common way to test the performance of the model in cybersecurity research~\cite{zhauniarovich_survey_2018}. Our final model achieved 80. 4\% precision with an AUC of 0.636, demonstrating a good predictive capacity to identify high-risk domains.
\end{itemize}

Together, these methods help us understand whether simple registration details available at the time of domain creation can act as early warning signs to spot risky toll scam domains before they start causing harm.

%% file: results.tex
\section{Results}

\subsection{RQ1: Infrastructure Patterns Distinguishing Suspended from Active Toll Scam Domains}

To investigate whether suspended toll scam domains exhibit distinct infrastructure characteristics, we tested whether they are disproportionately concentrated within a small number of ASNs and demonstrate greater IP address clustering compared to active domains. We analyze ASN distribution patterns to identify hosting providers with unusually high concentrations of suspended domains.

\subsubsection{ASN Distribution and Risk Concentration}

We examined the relationship between domain status and ASN distribution using chi-square analysis to test whether suspended domains concentrate on specific ASNs more than active domains.

The results revealed a highly significant association ($\chi^2 = 12{,}695.171$, $df = 67$, $p < .001$), indicating systematic differ- ences in ASN usage between suspended and active domains. This finding supports research documenting systematic abuse patterns in hosting infrastructure~\cite{coull_understanding_2012,konte_dynamics_2009}.

\begin{figure}[h!]
\centerline{\includegraphics[scale=0.40]{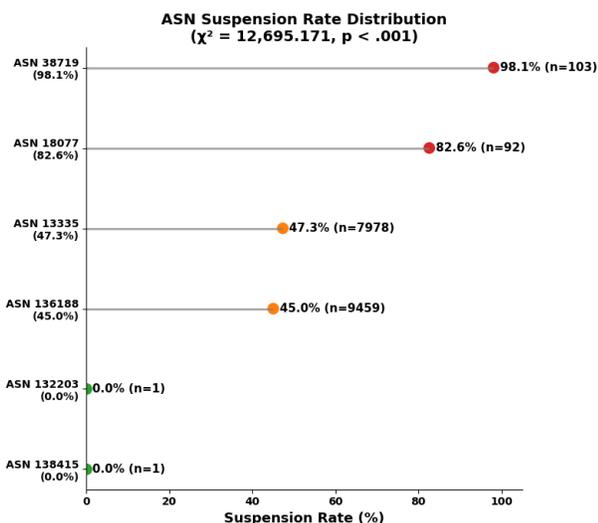}}

\caption{Distribution of domain suspension rates across ASNs. Suspension rates range from 0.0\% to 98.1\% across hosting providers, with sample sizes indicated.}

\label{fig:asn_risk_stratification}
\end{figure}

Crosstabulation analysis identified clear concentration pat- terns consistent with research on malicious hosting infras- tructure dynamics~\cite{konte_dynamics_2009}. As shown in Figure~\ref{fig:asn_risk_stratification}, certain ASNs exhibited disproportionately high suspension rates:

\begin{itemize}
\item ASN 38719: 98.1\% suspension rate ($n=103$)
\item ASN 18077: 82.6\% suspension rate ($n=92$)
\item ASN 13335: 47.3\% suspension rate ($n=7,978$)
\item ASN 136188: 45.0\% suspension rate ($n=9,459$)
\end{itemize}

In contrast, some ASNs maintained lower suspension rates, including ASN 132203 and ASN 138415 (both 0.0\% suspension rate, $n=1$ each), though these findings are limited by small sample sizes.

These suspension rate patterns demonstrate the systematic nature of infrastructure selection in malicious domain operations, supporting infrastructure-based risk assessment approaches documented in proactive detection literature~\cite{hao_predator_2016}. The concentration of suspended domains on specific ASNs aligns with research showing that certain hosting providers are systematically exploited for malicious activities~\cite{konte_dynamics_2009}.

\subsubsection{IP Address Clustering Analysis}

Building on our ASN findings, we assessed infrastructure concentration by computing \texttt{ip\_domain\_count} (the number of domains sharing each unique IP address) and comparing clustering patterns between suspended and active domains using independent-samples $t$-tests.

\begin{figure}[h!]
\centerline{\includegraphics[scale=0.42]{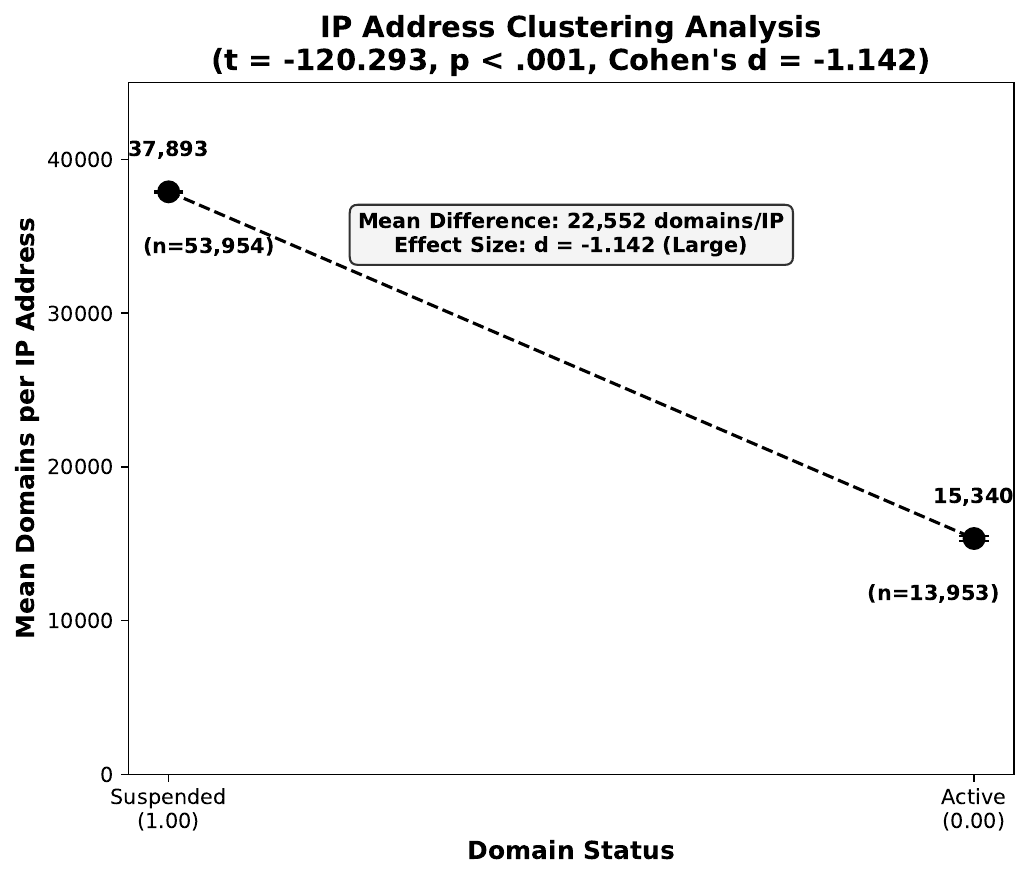}}

\caption{Comparison of IP address clustering between suspended and active toll scam domains. Suspended domains show significantly higher clustering compared to active domains.}
\label{fig:ip_clustering_comparison}
\end{figure}

The analysis revealed a significant difference ($t = -120.293$, $df = 67,905$, $p < .001$) with substantial practical significance. Figure~\ref{fig:ip_clustering_comparison} illustrates that suspended domains exhibited markedly higher IP clustering ($M = 37,892.58$, $SD = 19,065.77$) compared to active domains ($M = 15,340.09$, $SD = 22,154.05$), representing a mean difference of 22,552 domains per IP address (95\% CI: 22,185.035 to 22,919.954).

The large effect size (Cohen's $d = -1.142$) indicates that suspended domains host approximately 2.5 times more domains per IP address than active domains. This clustering pattern suggests either deliberate infrastructure concentration strategies or shared use of high-density hosting arrangements that facilitate operational efficiency but may also enable detection efforts.

To validate these findings against potential non-normality in the IP clustering data, we performed a Mann-Whitney U test, which confirmed the significant difference ($U = 197,252,753$, $Z = -107.099$, $p < .001$). Suspended domains showed significantly higher mean ranks (37,274.56) compared to active domains (21,113.94), providing robust evidence for the IP clustering hypothesis even when accounting for data distribution characteristics.

\subsubsection{Infrastructure Pattern Implications}

The observed patterns align with established research on domain registration abuses~\cite{coull_understanding_2012} and scam hosting infrastructure dynamics~\cite{konte_dynamics_2009}. The systematic concentration of suspended domains on specific ASNs suggests that certain hosting providers either lack adequate abuse monitoring capabilities or are deliberately exploited by malicious actors. Similarly, the extreme IP clustering among suspended domains indicates operational strategies that prioritize resource efficiency over detection avoidance.

These infrastructure characteristics provide measurable signals that distinguish suspended from active toll scam domains, supporting the development of proactive detection systems as demonstrated in domain abuse research~\cite{hao_predator_2016}. The systematic nature of these patterns suggests they reflect deliberate operational choices rather than random distribution.

Our analysis reveals distinct infrastructure patterns that distinguish suspended from active toll scam domains across two key dimensions:

\begin{itemize}
\item \textbf{ASN Concentration}: Suspended toll scam domains demonstrate significant concentration on specific ASNs, with suspension rates varying dramatically from 0.0\% to 98.1\% across hosting providers ($\chi^2 = 12,695.171$, $p < .001$)
\item \textbf{IP Clustering}: Suspended domains exhibit substantially higher IP address clustering, hosting approximately 2.5 times more domains per IP address than active domains ($t = -120.293$, $p < .001$, Cohen's $d = -1.142$)
\end{itemize}

Our analysis reveals distinct infrastructure patterns that distinguish suspended from active toll scam domains across two key dimensions:

\begin{itemize}
\item \textbf{ASN Concentration}: Suspended toll scam domains demonstrate significant concentration on specific ASNs, with suspension rates varying dramatically from 0.0\% to 98.1\% across hosting providers ($\chi^2 = 12,695.171$, $p < .001$)
\item \textbf{IP Clustering}: Suspended domains exhibit substantially higher IP address clustering, hosting approximately 2.5 times more domains per IP address than active domains ($t = -120.293$, $p < .001$, Cohen's $d = -1.142$)
\end{itemize}

These infrastructure patterns provide empirical evidence for systematic differences in hosting infrastructure selection and utilization between suspended and active toll scam domains. The observed concentration effects align with established research on malicious hosting infrastructure~\cite{konte_dynamics_2009} and suggest potential applications for infrastructure-based detection approaches~\cite{akiwate_risky_2021,hao_predator_2016}.

\subsection{RQ2: Toll Scam Registration Patterns Across TLDs and Registrars}

We analyzed registration patterns across 67,907 toll scam domains to characterize how toll scam operators distribute their activities across different TLDs and registrars.

\subsubsection{TLD Exploitation Patterns:}

\begin{figure}[h!]
\centerline{\includegraphics[scale=0.32]{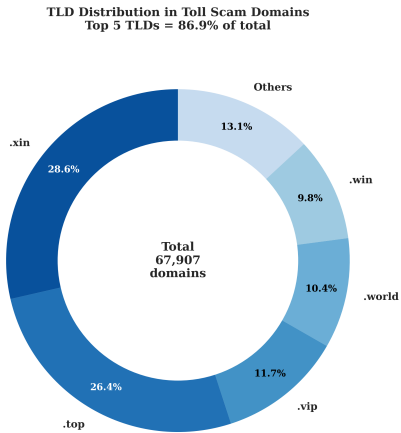}}

\caption{TLD concentration patterns showing extreme concentration in non-mainstream TLDs. The donut chart emphasizes the total scale of 67,907 domains while highlighting the systematic preference for non-mainstream TLDs. The visualization clearly demonstrates the deliberate avoidance of traditional, well-monitored domain spaces.}
\label{fig:tld_concentration}
\end{figure}

We conducted frequency analysis to identify which TLDs are most commonly used by toll scam operators, as this method reveals whether attackers systematically target specific domain spaces or register randomly across all available options~\cite{korczynski_cybercrime_2018}. Our analysis revealed extreme concentration in non-mainstream TLDs, with just five domains accounting for 86.9\% of all registrations: \texttt{.xin} (28.6\%), \texttt{.top} (26.4\%), \texttt{.vip} (11.7\%), \texttt{.world} (10.4\%), and \texttt{.win} (9.8\%). This distribution is clearly visualized in Figure~\ref{fig:tld_concentration} which looks very different from what we see in regular internet usage, where mainstream TLDs like \texttt{.com} typically make up the majority of domains. Surprisingly, \texttt{.com} represented only 1.3\% of our toll scam dataset, suggesting that scammers deliberately stay away from heavily monitored TLD spaces.

\subsubsection{Registrar Concentration Analysis}

\begin{figure}[h!]
    \centering 
    \includegraphics[scale=0.28]{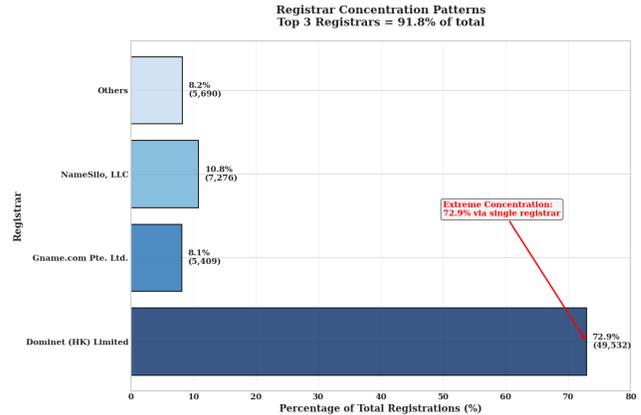}
    \caption{Registrar concentration patterns illustrating remarkable concentration through a horizontal bar chart. The annotations highlight how the top three registrars account for 91.8\% of all registrations. The visualization emphasizes the systematic targeting of registrars with minimal verification requirements.}
    \label{fig:registrar_concentration}
\end{figure}

After identifying concentration in TLD usage, we next examined whether similar patterns exist in registrar choices using the same frequency analysis approach to determine whether toll scam operators show preferences for specific registration services~\cite{akiwate_risky_2021}. The results demonstrated equally striking exploitation patterns, with Dominet (HK) Limited dominating 72.9\% of all toll scam registrations, followed by NameSilo (10.8\%) and Gname.com (8.1\%). Figure~\ref{fig:registrar_concentration} highlights this remarkable concentration. Together, these three registrars handled 91.8\% of all registrations. This concentration becomes particularly significant when compared to established registrars such as GoDaddy, Namecheap, and Google Domains, which each represented less than 0.1\% of registrations despite their significant market presence in legitimate domains.

\subsubsection{Infrastructure Clustering Validation}

To better understand the operational structure behind these concentration patterns, we applied the $K$ mean clustering analysis to identify natural groupings in the registration behavior, as this unsupervised method can reveal coordinated operational patterns without requiring prior assumptions about the structure of the campaign~\cite{vissers_exploring_2017}. This clustering approach revealed that almost all scam domains are tied to a single operational ecosystem, confirming strong systemic coordination. Specifically, cluster 1 encompassed 99. 9\% of domains characterized by registrations through Dominet and associated registrars using the five dominant TLDs, while cluster 2 contained only 36 domains (0.1\%) exclusively associated with NameSilo. ANOVA confirmed that the choice of the registrar was the key factor distinguishing different groups, while the selection of the TLD showed little variation between the clusters. The concentration was so extreme that traditional chi-square testing was not possible due to insufficient variation. Importantly, this inability to test variation is itself meaningful evidence that toll scam operators work systematically rather than randomly. Such a high concentration in both registrars and TLDs provides clear early warning signals to identify risky domains.

\subsection{RQ3: Temporal Patterns in Toll Scam Domain Registrations}

We analyze temporal registration patterns in 67,907 toll scam domains to identify when and how toll scam operators coordinate their domain registration activities.

\subsubsection{Monthly Registration Pattern Analysis}

We conducted time series analysis by aggregating domain registrations into monthly counts to identify whether registration activity occurs randomly or clusters during specific periods~\cite{hao_understanding_2013}. This approach helps answer our research question by revealing whether scammers register domains continuously or in coordinated bursts that suggest planned campaign launches~\cite{chatfield_analysis_2019}. ANOVA confirmed substantial differences in monthly registration volumes ($F = 2,241,965.784$, $df = 11, 66,017$, $p < .001$). February 2025 emerged as the dominant registration month with 24,342 domains---a striking 24-fold increase over months with minimal activity, where July 2024 and September 2024 each recorded only one domain registration.

\subsubsection{Temporal Coordination Validation.} To validate the clustering patterns and test for randomness, we applied the Runs Test, which statistically determines whether monthly registration sequences occur randomly or in systematic bursts~\cite{field_discovering_2018}. The test revealed only 3 runs across the entire dataset with a highly significant result ($Z = -260.576$, $p < .001$), providing strong statistical evidence against random distribution. Additionally, autocorrelation analysis revealed persistent temporal patterns supporting coordination evidence~\cite{chatfield_analysis_2019}. The autocorrelation function (ACF) showed high correlation coefficients across lags (1.000 at lag 1, decreasing to 0.997 at lag 24), while Box-Ljung statistics confirmed significant non-randomness (e.g., 1,625,697.690 at lag 24, $p < .001$). The partial autocorrelation function (PACF) showed strong influence from the immediate previous month (1.000 at lag 1) but minimal influence from earlier periods, suggesting that registration campaigns create short-term momentum followed by rapid decline.

\subsubsection{Statistical Validation.} Following these statistical tests, we used Tukey's HSD analysis to determine exactly which months were different from each other. This analysis confirmed that February's registration volume was significantly higher than all other months ($p < .001$), creating a completely separate statistical group. 

These findings together provide strong evidence that toll scam operators don't register domains randomly throughout the year. Instead, they clearly coordinate their activities, with February standing out as their preferred month for large-scale domain registration campaigns. This systematic approach shows that scammers are working in an organized way rather than acting independently.

For security teams, this extreme concentration during specific time periods gives them clear warning signs to watch for. By monitoring for unusual registration spikes, especially during high-risk months like February, detection systems can spot potential toll scam campaigns before they fully launch and start targeting victims.

\subsection{RQ4: Predictive Modeling of Toll Scam Domain Characteristics}

We developed predictive models to answer whether registration data can help identify which toll scam domains are likely to be suspended. This analysis provides the first comprehensive study of predicting toll scam domain behavior using infrastructure patterns.

\subsubsection{Creating Risk Categories}

We turned the patterns found in RQ2 into simple risk categories for our prediction model. Based on our earlier findings, we labeled five TLDs as ``high-risk'' because they contained most toll scam domains: .xin, .top, .vip, .world, and .win. Together, these five TLDs hosted 86.9\% of all toll scam domains in our dataset.

Similarly, we identified three ``high-risk'' registrars: Dominet HK Limited, NameSilo, and Gname.com. These three companies handled 91.8\% of all toll scam registrations, even though they normally handle only 3--6\% of legitimate domain registrations. This huge difference (91.8\% vs 3--6\%, $p < 0.000001$) shows that toll scam operators deliberately choose specific registrars, supporting findings from previous research on domain abuse patterns~\cite{akiwate_risky_2021,vissers_exploring_2017}.

\subsubsection{Model Results and Performance}

Of the 67,907 toll scam domains identified, our prediction model analyzed 65,977 domains with complete registration and status information, excluding entries with missing values.
The model achieved 80.4\% accuracy in predicting which domains would be suspended versus which would remain active. This result was statistically significant ($\chi^2 = 6,301.8$, $df = 2$, $p < 0.001$), meaning our model performs much better than random guessing. The Hosmer-Lemeshow test confirmed our model fits the data well ($\chi^2 = 0.948$, $p = 0.330$).

Both risk factors proved important for prediction. Domains using high-risk TLDs (.xin, .top, .vip, .world, .win) were 5.9 times more likely to be suspended than domains using mainstream TLDs like .com or .org (OR = 0.170, 95\% CI: 0.162--0.178, $p < 0.001$). Domains registered through high-risk registrars were 1.7 times more likely to be suspended (OR = 0.575, 95\% CI: 0.532--0.621, $p < 0.001$).

\subsubsection{Model Accuracy and Detection Capability}

\begin{figure}[h!]
\centerline{\includegraphics[scale=0.3]{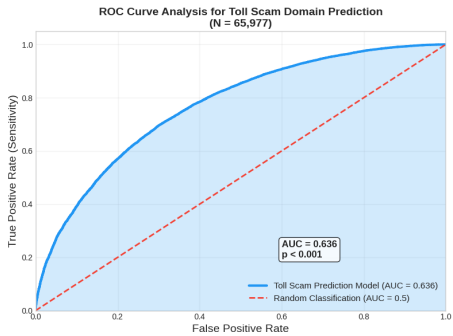}}

\caption{ROC curve analysis for toll scam domain suspension prediction. The model achieved AUC = 0.636 (p $<$ 0.001, N = 65{,}977) using TLD and registrar risk indicators, with 80.4\% accuracy and 92.3\% sensitivity}

\label{fig:roc_curve}
\end{figure}

We used ROC curve analysis to evaluate how well our model distinguishes between domains that will be suspended versus those that will stay active. The analysis produced an AUC (Area Under Curve) of 0.636 ($p < 0.001$).The ROC curve in Figure~\ref{fig:roc_curve} shows that an AUC of 0.636 means our model has a 63.6\% chance of correctly ranking a suspended domain as higher risk than an active domain. While not perfect, this performance level is considered useful for cybersecurity applications, where AUC scores above 0.6 are operationally valuable~\cite{zhauniarovich_survey_2018,hao_predator_2016}.

Most importantly, our model showed excellent sensitivity of 92.3\% for detecting suspended domains. This means the model correctly identified 92.3\% of domains that were actually suspended. This high detection rate is particularly valuable for cybersecurity because it means the model rarely misses high-risk domains that enforcement agencies should prioritize~\cite{lim_registration_2025}.

Our results show that simple registration information can effectively predict toll scam domain suspension risk. The 80.4\% accuracy using only TLD and registrar data means that security teams can assess domain risk immediately when domains are registered, before any attack occurs. This aligns with successful proactive detection systems developed for other types of domain abuse~\cite{hao_predator_2016}.

The clear odds ratios make the model practical to use: domains that combine high-risk TLDs with high-risk registrars can be automatically flagged for investigation. This allows security teams to focus their limited resources on the domains most likely to cause harm.

These findings establish that toll scam operators follow predictable registration patterns that can be detected early. Security systems can now identify high-risk toll scam domains at registration time, enabling prevention rather than just reaction after victims are harmed~\cite{coull_understanding_2012,korczynski_cybercrime_2018}. This represents a major step forward in toll scam detection, providing the first validated method for predicting toll scam domain behavior using registration data alone.

\subsection{Limitations and Future Directions}

This study offers valuable insights into the infrastructure patterns and predictive potential of toll scam domains based on a novel dataset of 67,907 domains. However, the following limitations should be acknowledged to contextualize the findings and inform future research directions.

Our analysis focuses primarily on domains registered in 2025, which provides excellent insight into current toll scam patterns. However, this means our findings might not capture how these patterns have changed over time or how they might evolve as scammers adapt to new detection methods. This temporal focus, while providing strong contemporary evidence, limits our ability to identify long-term trends in toll scam infrastructure evolution.

We identified problematic domains by looking at which ones were suspended by enforcement agencies, which is a standard approach used in cybersecurity research. This method gives us solid evidence of domains that were actually causing harm, since authorities only suspend domains after confirming malicious activity. However, we recognize that some dangerous domains might stay active longer than others before getting caught and suspended, which could affect how well our prediction model works in different situations. The timing of when authorities discover and shut down these domains can vary, meaning some toll scam operations might run longer before being detected.
\vspace{0.8em}

This dataset relies on toll scam domains identified and recorded by the State of Oklahoma, which uses targeted methods such as monitoring known hosting locations and passive DNS to discover domains. Consequently, the dataset may be biased toward the State's collection approach to identify high-risk domains, potentially missing patterns in toll scam domains that do not fit their methodology.

In the future, we will track toll scam patterns over several years to see if scammers keep using the same TLDs and registrars we found, or if they start switching to new ones as detection methods improve. We also aim to expand our work to compare scam versus non-scam domains, which would help us build better models that can clearly tell the difference between dangerous and safe websites.

%% file: discussion.tex
\section{Discussion}

\textbf{Systematic Infrastructure Exploitation Patterns:} Our analysis of 67,907 toll scam domains demonstrates systematic infrastructure selection strategies rather than opportunistic registration behavior. The extreme concentration observed indicates deliberate operational choices: 86.9\% of domains utilized five TLDs (\texttt{.xin}, \texttt{.top}, \texttt{.vip}, \texttt{.world}, \texttt{.win}) while 91.8\% registered through three providers. The minimal representation of mainstream TLDs such as \texttt{.com} (1.3\%) contrasts sharply with legitimate domain distribution patterns. This suggests sophisticated avoidance of heavily monitored registration spaces~\cite{korczynski_cybercrime_2018}.

\textbf{Temporal Coordination Evidence.} Registration patterns exhibit significant temporal clustering inconsistent with random distribution. February registrations exceeded low-activity months by a factor of 24,000. Statistical validation through Runs Test analysis revealed only three transitions between activity levels across the dataset ($Z = -260.576$, $p < .001$). This temporal concentration aligns with research on coordinated cybercriminal campaign launches~\cite{vissers_exploring_2017}. The timing suggests strategic coordination potentially targeting peak travel periods when toll road usage increases.

\textbf{Infrastructure Risk Differentiation: } Comparative analysis between suspended and active domains reveals systematic hosting infrastructure differences. Suspended domains demonstrate 2.5-fold higher IP address clustering compared to active domains. They also concentrate on ASNs with elevated suspension rates, ranging from 0.0\% to 98.1\% across providers. These patterns suggest either deliberate exploitation of permissive hosting environments or differential abuse monitoring capabilities across providers~\cite{akiwate_risky_2021}.

\textbf{Predictive Detection Implications:} The systematic nature of infrastructure selection enables proactive risk assessment using registration metadata. Logistic regression modeling achieved 80.4\% accuracy and 92.3\% sensitivity for identifying high-risk domains. This performance aligns with established benchmarks for operationally viable cybersecurity detection systems~\cite{zhauniarovich_survey_2018}.

\textbf{Theoretical and Methodological Contributions:} These findings challenge assumptions regarding cybercriminal operational coordination. They suggest centralized organizational structures underlying toll scam campaigns. The extreme concentration required methodological adaptations. Insufficient variation prevented traditional chi-square analysis, but this constraint itself constitutes evidence of systematic exploitation patterns. This necessitated analytical approaches specifically suited to highly skewed criminal behavior data, contributing methodological insights for cybersecurity research.

\textbf{Operational Security Implications:} The infrastructure concentration patterns we found give security teams practical information they can use to build better automated detection systems. Since toll scam operators consistently choose the same types of TLDs and registrars, security systems can watch for these specific combinations and alert teams when they appear. When we combine this with the timing patterns we discovered—especially the February spike—detection systems can be set to pay extra attention during high-risk months when toll scam activity typically increases.

However, our findings also highlight an important challenge for the future. The highly organized coordination we observed suggests that toll scam operators might adapt their strategies as security systems get better at detecting their current patterns. This means security teams will need monitoring approaches that can change and evolve along with criminal tactics, rather than relying on fixed detection rules that might become outdated as scammers learn to avoid them.

%% file: conclusion.tex
\section{Conclusion}
\label{sec:conclusion}

This study analyzed 67,907 toll scam domains to understand how cybercriminals choose and use domain registration infrastructure, and whether we can detect these patterns early to prevent attacks. Our findings reveal that toll scam operations are far more organized and systematic than previously understood.

We first examined what makes suspended domains different from active ones and found clear patterns in their hosting choices. Suspended domains cluster much more heavily on certain types of hosting infrastructure, hosting 2.5 times more domains per IP address than active ones. They also concentrate on specific hosting providers, with suspension rates ranging from 0\% to 98\% across providers. This suggests that scammers either target hosting services with weaker monitoring or that some providers are more vulnerable to exploitation.

In terms of registration behavior, we discovered extreme concentration rather than random selection. Just five TLDs (\texttt{.xin}, \texttt{.top}, \texttt{.vip}, \texttt{.world}, \texttt{.win}) account for nearly 87\% of all toll scam registrations, while three registrars handle over 91\% of these domains. Mainstream domains like \texttt{.com} constitute only 1.3\% of toll scams, showing that criminals deliberately avoid well-monitored spaces.

We also uncovered extreme temporal clustering in registration activity, with February showing a dramatic surge compared to quieter months. Statistical tests confirmed that this pattern is not random but reflects highly coordinated campaign launches, where large batches of domains are registered simultaneously to support scalable attacks.

Crucially, these systematic patterns enable early detection. We trained a prediction model using only registration metadata available at domain creation time to identify high-risk toll scam domains. The model achieved 80.4\% accuracy and 92.3\% sensitivity for predicting which domains would eventually be suspended. Domains using high-risk TLDs were nearly 6 times more likely to be suspended, while those registered through high-risk providers were 1.7 times more likely to face enforcement action.

Overall, this study shows that toll scam domains follow highly predictable patterns in where, how, and when they are registered. These insights can inform the development of early-warning systems that proactively flag risky domains before they inflict harm. In future work, we plan to investigate whether these patterns persist over time and explore whether similar methods can be extended to other forms of cybercrime.

%% file: acknowledgment.tex
\section{Acknowledgment}
The authors would like to express their sincere gratitude to Josh Swenson from the State of Oklahoma for generously providing the dataset that made this research possible. The authors would also like to thank the anonymous reviewers of eCrime 2025 for their valuable feedback, which helped us improve the camera-ready version of our paper.

%% file: main.bbl
\begin{thebibliography}{10}
\providecommand{\url}[1]{#1}
\csname url@samestyle\endcsname
\providecommand{\newblock}{\relax}
\providecommand{\bibinfo}[2]{#2}
\providecommand{\BIBentrySTDinterwordspacing}{\spaceskip=0pt\relax}
\providecommand{\BIBentryALTinterwordstretchfactor}{4}
\providecommand{\BIBentryALTinterwordspacing}{\spaceskip=\fontdimen2\font plus
\BIBentryALTinterwordstretchfactor\fontdimen3\font minus \fontdimen4\font\relax}
\providecommand{\BIBforeignlanguage}[2]{{%
\expandafter\ifx\csname l@#1\endcsname\relax
\typeout{** WARNING: IEEEtran.bst: No hyphenation pattern has been}%
\typeout{** loaded for the language `#1'. Using the pattern for}%
\typeout{** the default language instead.}%
\else
\language=\csname l@#1\endcsname
\fi
#2}}
\providecommand{\BIBdecl}{\relax}
\BIBdecl

\bibitem{mineo2024youd}
\BIBentryALTinterwordspacing
L.~Mineo, ``You’d never fall for an online scam, right?'' \emph{Harvard Gazette}, Sep. 2024. [Online]. Available: \url{https://news.harvard.edu/gazette/story/2024/09/youd-never-fall-for-an-online-scam-right/}
\BIBentrySTDinterwordspacing

\bibitem{rayo2025got}
\BIBentryALTinterwordspacing
A.~Rayo, ``Got a text about unpaid tolls? {It’s} probably a scam,'' Consumer Alert, Federal Trade Commission, Jan. 2025. [Online]. Available: \url{https://consumer.ftc.gov/consumer-alerts/2025/01/got-text-about-unpaid-tolls-its-probably-scam}
\BIBentrySTDinterwordspacing

\bibitem{IC3-2024-PSA240412}
\BIBentryALTinterwordspacing
{Internet Crime Complaint Center (IC3)}, ``Smishing {Scam} {Regarding} {Debt} for {Road} {Toll} {Services},'' Public Service Announcement, United States, Apr. 2024. [Online]. Available: \url{https://www.ic3.gov/PSA/2024/PSA240412}
\BIBentrySTDinterwordspacing

\bibitem{fbi_ic3_annualreport2024}
\BIBentryALTinterwordspacing
{Internet Crime Complaint Center (IC3)}, ``Internet {Crime} {Report} 2024,'' Federal Bureau of Investigation, Tech. Rep. IC3-2024, Apr. 2025. [Online]. Available: \url{https://www.ic3.gov/AnnualReport/Reports/2024_IC3Report.pdf}
\BIBentrySTDinterwordspacing

\bibitem{viu2025cybercrime}
\BIBentryALTinterwordspacing
{Vancouver Island University News}, ``Cybercrime's hidden toll: How online scams impact mental health,'' News release, Vancouver Island University, Mar. 2025. [Online]. Available: \url{https://news.viu.ca/cybercrimes-hidden-toll-how-online-scams-impact-mental-health}
\BIBentrySTDinterwordspacing

\bibitem{bidgoli2017hello}
M.~Bidgoli and J.~Grossklags, ``“hello. this is the irs calling.”: A case study on scams, extortion, impersonation, and phone spoofing,'' in \emph{2017 APWG Symposium on Electronic Crime Research (eCrime)}.\hskip 1em plus 0.5em minus 0.4em\relax IEEE, 2017, pp. 57--69.

\bibitem{Gibson2025UnpaidTollScam}
\BIBentryALTinterwordspacing
K.~Gibson and A.~M.~D. Lee, ``Unpaid toll texts sent nationwide are part of a con, federal and state officials warn,'' \emph{CBS News}, Feb. 2025. [Online]. Available: \url{https://www.cbsnews.com/news/unpaid-toll-text-scam/}
\BIBentrySTDinterwordspacing

\bibitem{hable2025phishing}
\BIBentryALTinterwordspacing
F.~Hable, N.-B. Schirrmacher, and B.~Hooff, ``Phishing {{Attacks}} in {{Context}}: {{Organizational Factors Shaping Phishing Susceptibility}},'' \emph{AMCIS 2025 Proceedings}, 2025. [Online]. Available: \url{https://aisel.aisnet.org/amcis2025/sig_sec/sig_sec/26}
\BIBentrySTDinterwordspacing

\bibitem{hao_predator_2016}
\BIBentryALTinterwordspacing
S.~Hao, A.~Kantchelian, B.~Miller, V.~Paxson, and N.~Feamster, ``{PREDATOR}: {Proactive} {Recognition} and {Elimination} of {Domain} {Abuse} at {Time}-{Of}-{Registration},'' in \emph{Proceedings of the 2016 {ACM} {SIGSAC} {Conference} on {Computer} and {Communications} {Security}}, ser. {CCS} '16.\hskip 1em plus 0.5em minus 0.4em\relax New York, NY, USA: Association for Computing Machinery, Oct. 2016, pp. 1568--1579. [Online]. Available: \url{https://dl.acm.org/doi/10.1145/2976749.2978317}
\BIBentrySTDinterwordspacing

\bibitem{kintis_hiding_2017}
\BIBentryALTinterwordspacing
P.~Kintis, N.~Miramirkhani, C.~Lever, Y.~Chen, R.~Romero-Gómez, N.~Pitropakis, N.~Nikiforakis, and M.~Antonakakis, ``Hiding in {Plain} {Sight}: {A} {Longitudinal} {Study} of {Combosquatting} {Abuse},'' in \emph{Proceedings of the 2017 {ACM} {SIGSAC} {Conference} on {Computer} and {Communications} {Security}}, ser. {CCS} '17.\hskip 1em plus 0.5em minus 0.4em\relax New York, NY, USA: Association for Computing Machinery, Oct. 2017, pp. 569--586. [Online]. Available: \url{https://dl.acm.org/doi/10.1145/3133956.3134002}
\BIBentrySTDinterwordspacing

\bibitem{coull_understanding_2012}
\BIBentryALTinterwordspacing
S.~E. Coull, A.~M. White, T.-F. Yen, F.~Monrose, and M.~K. Reiter, ``\BIBforeignlanguage{en}{Understanding domain registration abuses},'' \emph{\BIBforeignlanguage{en}{Computers \& Security}}, vol.~31, no.~7, pp. 806--815, Oct. 2012. [Online]. Available: \url{https://linkinghub.elsevier.com/retrieve/pii/S016740481200082X}
\BIBentrySTDinterwordspacing

\bibitem{agten2015seven}
P.~Agten, W.~Joosen, F.~Piessens, and N.~Nikiforakis, ``Seven months' worth of mistakes: A longitudinal study of typosquatting abuse,'' in \emph{Proceedings of the 22nd Network and Distributed System Security Symposium (NDSS 2015)}.\hskip 1em plus 0.5em minus 0.4em\relax Internet Society, 2015.

\bibitem{lim_registration_2025}
\BIBentryALTinterwordspacing
K.~Lim, R.~Sommese, M.~Jonker, R.~Mok, k.~claffy, and D.~Kim, ``Registration, {Detection}, and {Deregistration}: {Analyzing} {DNS} {Abuse} for {Phishing} {Attacks},'' Apr. 2025, arXiv:2502.09549 [cs]. [Online]. Available: \url{http://arxiv.org/abs/2502.09549}
\BIBentrySTDinterwordspacing

\bibitem{ICANNReports20241224}
\BIBentryALTinterwordspacing
{ICANN}, ``Icann {Contractual} {Compliance’s} enforcement of {DNS} {Abuse} requirements for {December} 2024,'' ICANN, Tech. Rep., 2024. [Online]. Available: \url{https://compliance-reports.icann.org/dnsabuse/dashboard/2024/1224/dns-abuse-report.html}
\BIBentrySTDinterwordspacing

\bibitem{alkhalil_phishing_2021}
\BIBentryALTinterwordspacing
Z.~Alkhalil, C.~Hewage, L.~Nawaf, and I.~Khan, ``Phishing {Attacks}: {A} {Recent} {Comprehensive} {Study} and a {New} {Anatomy},'' \emph{Frontiers in Computer Science}, vol.~3, 2021. [Online]. Available: \url{https://www.frontiersin.org/articles/10.3389/fcomp.2021.563060}
\BIBentrySTDinterwordspacing

\bibitem{fcc_how_2025}
\BIBentryALTinterwordspacing
{Federal Communications Commission}, ``How to {Spot} and {Avoid} {Toll} {Road} {Payment} {Scam} {Texts},'' retrieved 2025‑07‑23. [Online]. Available: \url{https://www.fcc.gov/consumer-governmental-affairs/how-spot-and-avoid-toll-road-payment-scam-texts}
\BIBentrySTDinterwordspacing

\bibitem{phishing_and_Countermeasures_2006}
\BIBentryALTinterwordspacing
eds M.~Jakobsson and S.~Myers, \emph{Phishing Attacks: Information Flow and Chokepoints}.\hskip 1em plus 0.5em minus 0.4em\relax John Wiley \& Sons, Ltd, 2006, ch.~2, pp. 31--63. [Online]. Available: \url{https://onlinelibrary.wiley.com/doi/abs/10.1002/9780470086100.ch2}
\BIBentrySTDinterwordspacing

\bibitem{dhamija_why_2006}
\BIBentryALTinterwordspacing
R.~Dhamija, J.~D. Tygar, and M.~Hearst, ``Why phishing works,'' in \emph{Proceedings of the {SIGCHI} {Conference} on {Human} {Factors} in {Computing} {Systems}}, ser. {CHI} '06.\hskip 1em plus 0.5em minus 0.4em\relax New York, NY, USA: Association for Computing Machinery, Apr. 2006, pp. 581--590. [Online]. Available: \url{https://dl.acm.org/doi/10.1145/1124772.1124861}
\BIBentrySTDinterwordspacing

\bibitem{rahman2023users}
M.~L. Rahman, D.~Timko, H.~Wali, and A.~Neupane, ``Users really do respond to smishing,'' in \emph{Proceedings of the Thirteenth ACM Conference on Data and Application Security and Privacy}, 2023, pp. 49--60.

\bibitem{jagatic_social_2007}
\BIBentryALTinterwordspacing
T.~N. Jagatic, N.~A. Johnson, M.~Jakobsson, and F.~Menczer, ``Social phishing,'' \emph{Commun. ACM}, vol.~50, no.~10, pp. 94--100, Oct. 2007. [Online]. Available: \url{https://dl.acm.org/doi/10.1145/1290958.1290968}
\BIBentrySTDinterwordspacing

\bibitem{oest_sunrise_2020}
\BIBentryALTinterwordspacing
A.~Oest, P.~Zhang, B.~Wardman, E.~Nunes, J.~Burgis, A.~Zand, K.~Thomas, A.~Doupé, and G.-J. Ahn, ``\BIBforeignlanguage{en}{Sunrise to {Sunset}: {Analyzing} the {End}-to-end {Life} {Cycle} and {Effectiveness} of {Phishing} {Attacks} at {Scale}},'' in \emph{\BIBforeignlanguage{en}{29th USENIX Security Symposium (USENIX Security '20)}}, 2020, pp. 361--377. [Online]. Available: \url{https://www.usenix.org/conference/usenixsecurity20/presentation/oest-sunrise}
\BIBentrySTDinterwordspacing

\bibitem{korczynski_cybercrime_2018}
\BIBentryALTinterwordspacing
M.~Korczynski, M.~Wullink, S.~Tajalizadehkhoob, G.~C.~M. Moura, A.~Noroozian, D.~Bagley, and C.~Hesselman, ``Cybercrime {After} the {Sunrise}: {A} {Statistical} {Analysis} of {DNS} {Abuse} in {New} {gTLDs},'' in \emph{Proceedings of the 2018 on {Asia} {Conference} on {Computer} and {Communications} {Security}}, ser. {ASIACCS} '18.\hskip 1em plus 0.5em minus 0.4em\relax New York, NY, USA: Association for Computing Machinery, May 2018, pp. 609--623. [Online]. Available: \url{https://dl.acm.org/doi/10.1145/3196494.3196548}
\BIBentrySTDinterwordspacing

\bibitem{akiwate_risky_2021}
\BIBentryALTinterwordspacing
G.~Akiwate, S.~Savage, G.~M. Voelker, and K.~C. Claffy, ``Risky {BIZness}: risks derived from registrar name management,'' in \emph{Proceedings of the 21st {ACM} {Internet} {Measurement} {Conference}}, ser. {IMC} '21.\hskip 1em plus 0.5em minus 0.4em\relax New York, NY, USA: Association for Computing Machinery, Nov. 2021, pp. 673--686. [Online]. Available: \url{https://dl.acm.org/doi/10.1145/3487552.3487816}
\BIBentrySTDinterwordspacing

\bibitem{konte_dynamics_2009}
M.~Konte, N.~Feamster, and J.~Jung, ``\BIBforeignlanguage{en}{Dynamics of {Online} {Scam} {Hosting} {Infrastructure}},'' in \emph{\BIBforeignlanguage{en}{Passive and {Active} {Network} {Measurement}}}, S.~B. Moon, R.~Teixeira, and S.~Uhlig, Eds.\hskip 1em plus 0.5em minus 0.4em\relax Berlin, Heidelberg: Springer, 2009, pp. 219--228.

\bibitem{cartron_dangers_2025}
\BIBentryALTinterwordspacing
N.~Cartron, ``\BIBforeignlanguage{en}{The {Dangers} of {DNS} {Hijacking}},'' Jan. 2025. [Online]. Available: \url{https://www.f5.com/labs/articles/threat-intelligence/the-dangers-of-dns-hijacking}
\BIBentrySTDinterwordspacing

\bibitem{holz_measuring_2008}
\BIBentryALTinterwordspacing
T.~Holz, C.~Gorecki, K.~Rieck, and F.~C. Freiling, ``\BIBforeignlanguage{en-US}{Measuring and {Detecting} {Fast}-{Flux} {Service} {Networks}},'' \emph{\BIBforeignlanguage{en-US}{NDSS Symposium}}, 2008-02-08. [Online]. Available: \url{https://www.ndss-symposium.org/ndss2008/measuring-and-detecting-fast-flux-service-networks/}
\BIBentrySTDinterwordspacing

\bibitem{caglayan_real-time_2009}
\BIBentryALTinterwordspacing
A.~Caglayan, M.~Toothaker, D.~Drapeau, D.~Burke, and G.~Eaton, ``Real-{Time} {Detection} of {Fast} {Flux} {Service} {Networks},'' in \emph{2009 {Cybersecurity} {Applications} \& {Technology} {Conference} for {Homeland} {Security}}, Mar. 2009, pp. 285--292. [Online]. Available: \url{https://ieeexplore.ieee.org/document/4804457}
\BIBentrySTDinterwordspacing

\bibitem{dingledine_evil_2009}
\BIBentryALTinterwordspacing
T.~Moore and R.~Clayton, ``\BIBforeignlanguage{en}{Evil {Searching}: {Compromise} and {Recompromise} of {Internet} {Hosts} for {Phishing}},'' in \emph{\BIBforeignlanguage{en}{Financial {Cryptography} and {Data} {Security}}}, R.~Dingledine and P.~Golle, Eds.\hskip 1em plus 0.5em minus 0.4em\relax Berlin, Heidelberg: Springer Berlin Heidelberg, 2009, vol. 5628, pp. 256--272, series Title: Lecture Notes in Computer Science. [Online]. Available: \url{http://link.springer.com/10.1007/978-3-642-03549-4_16}
\BIBentrySTDinterwordspacing

\bibitem{maroofi_comar_2020}
\BIBentryALTinterwordspacing
S.~Maroofi, M.~Korczyński, C.~Hesselman, B.~Ampeau, and A.~Duda, ``{COMAR}: {Classification} of {Compromised} versus {Maliciously} {Registered} {Domains},'' in \emph{2020 {IEEE} {European} {Symposium} on {Security} and {Privacy} ({EuroS}\&{P})}, Sep. 2020, pp. 607--623. [Online]. Available: \url{https://ieeexplore.ieee.org/document/9230367}
\BIBentrySTDinterwordspacing

\bibitem{erfan_owned_2024}
\BIBentryALTinterwordspacing
M.~Erfan, P.~Branco, and G.-V. Jourdan, ``Owned, {Pwned} or {Rented}: {Whose} {Domain} {Is} {It}?'' in \emph{2024 {APWG} {Symposium} on {Electronic} {Crime} {Research} ({eCrime})}, Sep. 2024, pp. 14--26. [Online]. Available: \url{https://ieeexplore.ieee.org/document/10896052}
\BIBentrySTDinterwordspacing

\bibitem{nahapetyan_sms_2024}
\BIBentryALTinterwordspacing
A.~Nahapetyan, S.~Prasad, K.~Childs, A.~Oest, Y.~Ladwig, A.~Kapravelos, and B.~Reaves, ``On {SMS} {Phishing} {Tactics} and {Infrastructure},'' in \emph{2024 {IEEE} {Symposium} on {Security} and {Privacy} ({SP})}, May 2024, pp. 1--16, iSSN: 2375-1207. [Online]. Available: \url{https://ieeexplore.ieee.org/document/10646609/}
\BIBentrySTDinterwordspacing

\bibitem{timko2023commercial}
\BIBentryALTinterwordspacing
D.~Timko and M.~L. Rahman, ``{Commercial Anti-Smishing Tools and Their Comparative Effectiveness Against Modern Threats},'' in \emph{Proceedings of the 16th ACM Conference on Security and Privacy in Wireless and Mobile Networks}, ser. WiSec '23.\hskip 1em plus 0.5em minus 0.4em\relax New York, NY, USA: Association for Computing Machinery, 2023, p. 1–12. [Online]. Available: \url{https://doi.org/10.1145/3558482.3590173}
\BIBentrySTDinterwordspacing

\bibitem{timko2024smishing}
D.~Timko and M.~L. Rahman, ``Smishing dataset i: Phishing sms dataset from smishtank. com,'' in \emph{Proceedings of the Fourteenth ACM Conference on Data and Application Security and Privacy}, 2024, pp. 289--294.

\bibitem{smishtank2025}
D.~Timko and M.~L. Rahman, ``Smishtank - smishing analysis tool,'' \url{https://smishtank.com/}, 2025, accessed: September 23, 2025.

\bibitem{moura_characterizing_2024}
\BIBentryALTinterwordspacing
G.~C.~M. Moura, T.~Daniels, M.~Bosteels, S.~Castro, M.~Müller, T.~Wabeke, T.~van~den Hout, M.~Korczyński, and G.~Smaragdakis, ``Characterizing and {Mitigating} {Phishing} {Attacks} at {ccTLD} {Scale},'' in \emph{Proceedings of the 2024 on {ACM} {SIGSAC} {Conference} on {Computer} and {Communications} {Security}}, ser. {CCS} '24.\hskip 1em plus 0.5em minus 0.4em\relax New York, NY, USA: Association for Computing Machinery, Dec. 2024, pp. 2147--2161. [Online]. Available: \url{https://dl.acm.org/doi/10.1145/3658644.3690192}
\BIBentrySTDinterwordspacing

\bibitem{ICANNReports20171975}
\BIBentryALTinterwordspacing
{ICANN}, ``{{ICANN}} {Reports} 2017,'' ICANN, Tech. Rep., 2017. [Online]. Available: \url{https://www.icann.org/en/system/files/files/annual-report-2017-en.pdf}
\BIBentrySTDinterwordspacing

\bibitem{vissers_exploring_2017}
T.~Vissers, J.~Spooren, P.~Agten, D.~Jumpertz, P.~Janssen, M.~Van~Wesemael, F.~Piessens, W.~Joosen, and L.~Desmet, ``\BIBforeignlanguage{en}{Exploring the {Ecosystem} of {Malicious} {Domain} {Registrations} in the .eu {TLD}},'' in \emph{\BIBforeignlanguage{en}{Research in {Attacks}, {Intrusions}, and {Defenses}}}, M.~Dacier, M.~Bailey, M.~Polychronakis, and M.~Antonakakis, Eds.\hskip 1em plus 0.5em minus 0.4em\relax Cham: Springer International Publishing, 2017, pp. 472--493.

\bibitem{tajalizadehkhoob_rotten_2018}
\BIBentryALTinterwordspacing
S.~Tajalizadehkhoob, R.~Böhme, C.~Gañán, M.~Korczyński, and M.~V. Eeten, ``Rotten {Apples} or {Bad} {Harvest}? {What} {We} {Are} {Measuring} {When} {We} {Are} {Measuring} {Abuse},'' \emph{ACM Trans. Internet Technol.}, vol.~18, no.~4, pp. 49:1--49:25, Aug. 2018. [Online]. Available: \url{https://dl.acm.org/doi/10.1145/3122985}
\BIBentrySTDinterwordspacing

\bibitem{hao_understanding_2013}
\BIBentryALTinterwordspacing
S.~Hao, M.~Thomas, V.~Paxson, N.~Feamster, C.~Kreibich, C.~Grier, and S.~Hollenbeck, ``Understanding the domain registration behavior of spammers,'' in \emph{Proceedings of the 2013 conference on {Internet} measurement conference}, ser. {IMC} '13.\hskip 1em plus 0.5em minus 0.4em\relax New York, NY, USA: Association for Computing Machinery, Oct. 2013, pp. 63--76. [Online]. Available: \url{https://dl.acm.org/doi/10.1145/2504730.2504753}
\BIBentrySTDinterwordspacing

\bibitem{agarwal_examining_2025}
\BIBentryALTinterwordspacing
S.~Agarwal and M.~Vasek, ``\BIBforeignlanguage{en}{Examining {Newly} {Registered} {Phishing} {Domains} at {Scale}},'' in \emph{\BIBforeignlanguage{en}{The 24th {Workshop} on the {Economics} of {Information} {Security}}}, Tokyo, Japan, 2025. [Online]. Available: \url{https://discovery.ucl.ac.uk/id/eprint/10209951/1/CDA_Domains___WEIS_25.pdf}
\BIBentrySTDinterwordspacing

\bibitem{hao_monitoring_2011}
\BIBentryALTinterwordspacing
S.~Hao, N.~Feamster, and R.~Pandrangi, ``Monitoring the initial {DNS} behavior of malicious domains,'' in \emph{Proceedings of the 2011 {ACM} {SIGCOMM} conference on {Internet} measurement conference}, ser. {IMC} '11.\hskip 1em plus 0.5em minus 0.4em\relax New York, NY, USA: Association for Computing Machinery, Nov. 2011, pp. 269--278. [Online]. Available: \url{https://dl.acm.org/doi/10.1145/2068816.2068842}
\BIBentrySTDinterwordspacing

\bibitem{zhauniarovich_survey_2018}
\BIBentryALTinterwordspacing
Y.~Zhauniarovich, I.~Khalil, T.~Yu, and M.~Dacier, ``A {Survey} on {Malicious} {Domains} {Detection} through {DNS} {Data} {Analysis},'' \emph{ACM Comput. Surv.}, vol.~51, no.~4, pp. 67:1--67:36, Jul. 2018. [Online]. Available: \url{https://dl.acm.org/doi/10.1145/3191329}
\BIBentrySTDinterwordspacing

\bibitem{moubayed_dns_2018}
\BIBentryALTinterwordspacing
A.~Moubayed, M.~Injadat, A.~Shami, and H.~Lutfiyya, ``{DNS} {Typo}-{Squatting} {Domain} {Detection}: {A} {Data} {Analytics} \& {Machine} {Learning} {Based} {Approach},'' in \emph{2018 {IEEE} {Global} {Communications} {Conference} ({GLOBECOM})}, Dec. 2018, pp. 1--7. [Online]. Available: \url{https://ieeexplore.ieee.org/document/8647679}
\BIBentrySTDinterwordspacing

\bibitem{shirazi_kn0w_2018}
\BIBentryALTinterwordspacing
H.~Shirazi, B.~Bezawada, and I.~Ray, ````{Kn0w} {Thy} {Doma1n} {Name}'': {Unbiased} {Phishing} {Detection} {Using} {Domain} {Name} {Based} {Features},'' in \emph{Proceedings of the 23nd {ACM} on {Symposium} on {Access} {Control} {Models} and {Technologies}}, ser. {SACMAT} '18.\hskip 1em plus 0.5em minus 0.4em\relax New York, NY, USA: Association for Computing Machinery, Jun. 2018, pp. 69--75. [Online]. Available: \url{https://dl.acm.org/doi/10.1145/3205977.3205992}
\BIBentrySTDinterwordspacing

\bibitem{prakash_phishnet_2010}
P.~Prakash, M.~Kumar, R.~R. Kompella, and M.~Gupta, ``Phishnet: predictive blacklisting to detect phishing attacks,'' in \emph{Proceedings of the 29th conference on {Information} communications}, ser. {INFOCOM}'10.\hskip 1em plus 0.5em minus 0.4em\relax San Diego, California, USA: IEEE Press, Mar. 2010, pp. 346--350.

\bibitem{field_discovering_2018}
A.~Field, \emph{Discovering Statistics Using {IBM} {SPSS} Statistics}, 5th~ed.\hskip 1em plus 0.5em minus 0.4em\relax London: SAGE Publications, 2018.

\bibitem{chatfield_analysis_2019}
C.~Chatfield and H.~Xing, \emph{The Analysis of Time Series: An Introduction with {R}}, 7th~ed.\hskip 1em plus 0.5em minus 0.4em\relax Boca Raton: CRC Press, 2019.

\end{thebibliography}
